\documentclass[twoside, epsf]{article}

\input ibvs2.sty
\input epsf.sty

\begin{document}

\IBVShead{5763}{10 April 2007}

\IBVStitle{SDSS J102146.44+234926.3: New WZ Sge-type dwarf nova}

\IBVSauth{Golovin, Alex$^{1, 2, 3}$; Ayani, Kazuya$^4$; Pavlenko, Elena P.$^5$; Krajci, Tom$^6$; Kuznyetsova, Yuliana$^{2,7}$; Henden, Arne$^8$; Krushevska, Victoria$^2$; Dvorak, Shawn$^9$; Sokolovsky, Kirill$^{10,11}$; Sergeeva, Tatyana P.$^2$; James, Robert$^{12}$; Crawford, Tim$^{13}$; Corp, Laurent$^{14}$;}

\IBVSinst{Kyiv National Taras Shevchenko University, Kyiv, UKRAINE
\\  \indent  \indent e-mail: astronom\_2003@mail.ru, astron@mao.kiev.ua} 

\IBVSinst{Main Astronomical Observatory of National Academy of
Science of Ukraine, Kyiv, UKRAINE}

\IBVSinst{Visiting astronomer of the Crimean Astrophysical
Observatory, Crimea, Nauchnyj, UKRAINE}

\IBVSinst{Bisei Astronomical Observatory, Ibara, Okayama, JAPAN}

\IBVSinst{Crimean Astrophysical Observatory, Crimea, Nauchnyj,
UKRAINE}

\IBVSinst{AAVSO, Cloudcroft, New Mexico, USA}

\IBVSinst{International Center of Astronomical and
Medico-Ecological Researches, Kyiv, UKRAINE}

\IBVSinst{AAVSO, Clinton B. Ford Astronomical Data and Research
Center, Cambridge, MA, USA}

\IBVSinst{Rolling Hills Observatory, Clermont, FL, USA}

\IBVSinst{Sternberg Astronomical Institute, Moscow State
University, Moscow, RUSSIA}

\IBVSinst{Astro Space Center of the Lebedev Physical Institute, Russian Academy of
Sciences, Moscow, RUSSIA}

\IBVSinst{AAVSO, Las Cruses, NM, USA}

\IBVSinst{AAVSO, Arch Cape Observatory, Arch Cape, OR, USA}

\IBVSinst{AAVSO, Rodez, FRANCE}

\IBVStyp{ WZ Sge }

\SIMBADobjAlias{SDSS J102146.44+234926.3}{USNO-B1.0 1138-0175054}

\IBVSkey{Stars: individual: SDSS J102146.44+234926.3 - stars: dwarf novae - stars: cataclysmic variables - stars: accretion, accretion discs - stars: spiral waves - stars: superhumps, double-humped - stars: echo outbursts, rebrightenings - methods: O-C - methods: periodogram - techniques: photometric - techniques: spectroscopic - techniques: photographic plates}

\IBVSabs{We report CCD photometry and spectroscopy during 2006}
\IBVSabs{outburst of the dwarf nova SDSS J102146.44+234926.3 (SDSS J1021).}
\IBVSabs{The photographic plates from the MAO, SAI and CrAO plate archives,}
\IBVSabs{which cover the position of the SDSS J1021, were inspected for the}
\IBVSabs{presence of previous outbursts. We also present the BVRcIc}
\IBVSabs{photometric calibration of 52 stars in SDSS J1021 vicinity, which}
\IBVSabs{have V-magnitude in the range of 11.21-17.23m and can serve}
\IBVSabs{as comparison stars. The large amplitude of the SDSS J1021}
\IBVSabs{outburst of 7m, superhumps with a period below the ''period}
\IBVSabs{gap'', rebrightening during the declining stage of superoutburst,}
\IBVSabs{rarity of outbursts and obtained spectrum allow to classify this}
\IBVSabs{object as a WZ Sge type dwarf nova.}

\begintext
The cataclysmic variable SDSS J102146.44+234926.3 (SDSS J1021
hereafter; $\alpha_{2000} = 10\hr 21\mm 46\fsec44; \delta_{2000} =
+23\deg 49\arcm 26\farcs3$) was discovered in outburst having a V
magnitude of $13\fmm9$ by Christensen on CCD images obtained in
the course of the Catalina Sky Survey on October 28.503 UT 2006.
In an archival image there is a star with $V \sim 21\mm$ at this
position (Christensen, 2006) and there is an object in the
database of the \emph{Sloan Digital Sky Survey} Data Release 5
(Adelman-McCarthy et al., 2007; SDSS DR5 hereafter) with the
following magnitudes, measured on January 17.455 UT, 2005: u =
20.83, g = 20.74, r = 20.63, i = 20.84, z = 20.45. In the
USNO-B1.0 catalog this object is listed as USNO-B1.0 1138-0175054
with magnitudes $B_{2mag} = 20.79$ and $R_{2mag}=20.35$. The large
amplitude and the blue color imply that the object could be a
dwarf nova of SU UMa or WZ Sge type (Waagen, 2006).

\begin{table}[!hb]
\begin{center}
{\normalsize Table 1. Log of observations} \vskip 3mm
\begin{tabular}{|cccccc|}
\hline
JD & Duration of & & & & \\
(mid of  & observational & Observatory & Telescope & CCD & Filter \\
obs. run) &  run [minutes] & & & & \\
\hline & & & & & \\
2454060.9 & 214 & Rolling Hills, FL, USA & Meade LX200-10 & SBIG ST-9 & V \\
2454061.0 & 158 & Cloudcroft, NM, USA & C-11 & SBIG ST-7 & none \\
2454062.0 & 259 & Cloudcroft, NM, USA & C-11 & SBIG ST-7 & none \\
2454062.9 & 288 & Cloudcroft, NM, USA & C-11 & SBIG ST-7 & none \\
2454063.6 & 115 & CrAO, UKRAINE & K-380 & SBIG ST-9 & R \\
2454064.6 & 222 & CrAO, UKRAINE & K-380 & SBIG ST-9 & R \\
2454066.7 & S.D.P. * & Pic du Midi, FRANCE & T-60 & Mx516 & None \\
2454067.6 & 90 & CrAO, UKRAINE & K-380 & Apogee 47p & R \\
2454067.9 & S.D.P. & Las Cruses, NM, USA & Meade LX200 & SBIG ST-7 & V \\
2454069.0 & S.D.P. & Arch Cape, USA & SCT-30 & SBIG ST-9 & V \\
2454069.0 & S.D.P. & Las Cruses, NM, USA & Meade LX200 & SBIG ST-7 & V \\
2454069.6 & 63 & CrAO, UKRAINE & K-380 & Apogee 47p & R \\
2454071.9 & S.D.P. & Las Cruses, NM, USA & Meade LX200 & SBIG ST-7 & V \\
2454072.9 & S.D.P. & Las Cruses, NM, USA & Meade LX200 & SBIG ST-7 & V \\
2454073.9 & S.D.P. & Las Cruses, NM, USA & Meade LX200 & SBIG ST-7 & V \\
2454074.9 & S.D.P. & Las Cruses, NM, USA & Meade LX200 & SBIG ST-7 & V \\
2454075.9 & S.D.P. & Las Cruses, NM, USA & Meade LX200 & SBIG ST-7 & V \\
2454166.8 & S.D.P. & Sonoita Observatory, USA & 0.35 m telescope & SBIG STL-1001XE & V \\
2454167.7 & S.D.P. & Sonoita Observatory, USA & 0.35 m telescope & SBIG STL-1001XE & V \\
\hline
\end{tabular}
\end{center}
* S.D.P. - Single Data Point
\end{table}

Fig.~1 (left) shows the $8\arcm \times 8\arcm$ image of the
SDSS J1021 vicinity, generated from SDSS DR5 Finding Chart Tool
(\emph{http://cas.sdss.org/astrodr5/en/tools/chart/chart.asp}).

\IBVSfig{6cm}{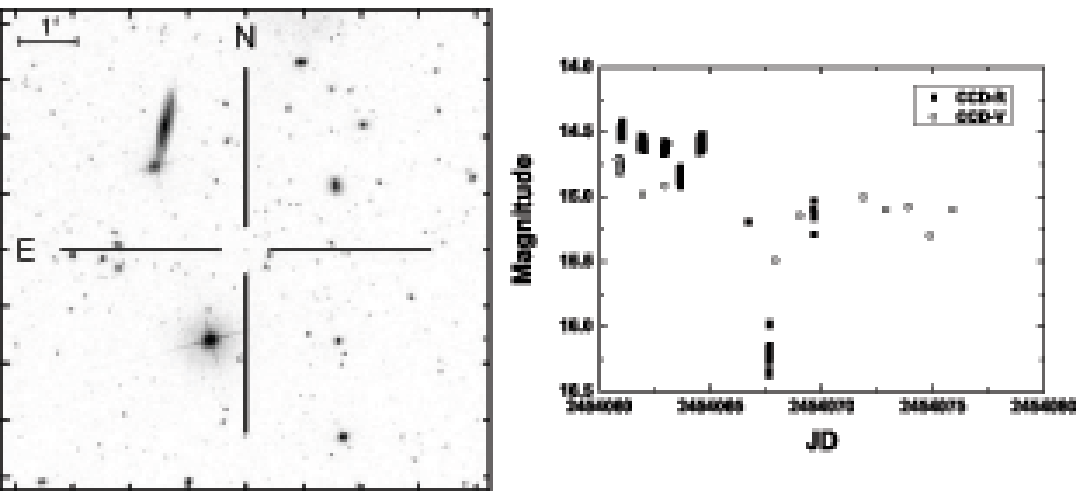}{\emph{Left:} SDSS image of the SDSS J1021
vicinity; \\ \emph{Right:} Light curve of SDSS J1021 during the
outburst;} \IBVSfigKey{5763-f1.eps}{SDSS J102146.44+234926.3}{light
curve and finding chart}

Time resolved CCD photometry has been carried out from different
sites by the authors since November 21, 2006 (the first night
after the discovery was reported) until 2006 December 06 (Data
available for download at {\tt http://www.aavso.org/data/download}
and from IBVS server; See Table~1 for log of observations). The
photometry was done in the V and $R_c$ bands as well as
unfiltered; this did not affect the following period analysis. The
error of a single measurement can be typically assumed to be $\pm
0\fmm02$. Fig.~1 (right) shows the overall light curve of the
object. Here we assume $m_R = m_{unfiltered}$. The light curve
could be divided into three parts, denoting the plateau stage, dip
and long-lasting echo-outburst (rebrightening).

Before carrying out Fourier analysis for the presence of
short-periodic signal in the light curve (superhumps), each
observer's data set was individually transformed to a uniform
zero-point by subtracting a linear fit from each night's
observations. This was done to remove the overall trend of the
outburst and to combine all observations into a single data set.

From the periodogram analysis (Fig.~2, left) the value of the
superhump period $P_{sh}$ = 0 \fday 05633 $\pm$ 0.00003 was
determined. Such a value is typical for the WZ Sge-type systems and is just
58.7 seconds shorter than $P_{sh}$ of another WZ Sge-like system:
ASAS 002511+1217.2 (Golovin et al., 2005).

The superhump light curve (with 15-point binning used) folded with
$0^d.05633$ period is shown on Fig.~2 (right). It is plotted for
two cycles for clarity. Only JD 2454061.0-2454063.6 data was
included. Note the $0\fmm1$ amplitude of variations and the
double-humped profile of the light curve. There remain many
questions concerning the nature of a double-humped superhumps in
the WZ Sge-type stars. The explanation of a double-humped light
curve could lie in a formation of a two-armed precessional spiral
density wave in the accretion disk (Osaki, 2003) or a one-armed
\textbf{optically thick} spiral wave, but with the occurrence of a
self-eclipse of the energy emitting source in the wave (Bisikalo,
2006).

Other theories concerning a double-peaked superhumps can be
found in Lasota et al. (1995), Osaki \& Meyer (2002), Kato (2002),
Patterson et al. (2002), Osaki \& Meyer (2003).

Applying the method of ''sliding parabolas'' (Marsakova \&
Andronov, 1996) we determined, when it was possible (JD 2454061.0
- 2454063.6), the times of maxima of superhumps (with mean $1
\sigma$ error of 0 \fday 0021) and calculated O-C residuals based
on founded period. The moments of superhump maximua are given in
Table~2. No period variations reaching the $3 \sigma$ level were
found during the time of observations. 

Another prominent feature of the SDSS J1021 light curve is the
echo-outburst (or \emph{rebrightening} - another term for this
event) that occurs during the declining stage of the
superoutburst. On Nov. 27/28 2006 (i.e. JD 2454067.61-2454067.68)
a rapid brightening with the rate of 0\fmm13 per hour was detected
at Crimean Astrophysical Observatory (Ukraine; CrAO hereafter),
that most probably was the early beginning of the echo-outburst.
Judging from our light curve, we conclude that rebrightening phase
lasted at least 8 days. Similar echo-outbursts are classified as
"type-A" echo-outburst according to classification system proposed
by Imada et al. (2006) as observed in the 2005 superoutburst of
TSS J022216.4+412259.9 and the 1995 superoutburst of AL Com (Imada
et al., 2006; Patterson et al., 1996).

Rebrightenings during the decline stage are observed in the WZ
Sge-type dwarf novae (as well as in some of the WZ Sge-type
candidate systems). However, their physical mechanism is still
poorly understood. In most cases, just one rebrightening
occurs (also observed sometimes in typical SU UMa systems), though
a series of rebrightenings are also possible, as it was manifested
by WZ Sge itself (12 rebrightenings), SDSS J0804 (11) and EG Cnc
(6) (Pavlenko et al., 2007). There are several competing theories
concerning what causes an echo-outburst(s) in such systems, though
all of them predict that the disk must be heated over the thermal
instability limit for a rebrightening to occur. See papers by
Patterson et al. (1998), Buat-Menard \& Hameury (2002), Schreiber
\& Gansicke (2001), Osaki, Meyer \& Meyer-Hofmeister (2001) and
Matthews et al. (2005) for a discussion of the physical reasons
for echo-outbursts.

Recent CCD-V photometry manifests that SDSS J1021 has a magnitude
of $19\fmm72 \pm 0.07$ and $19\fmm59 \pm 0.07$ as of 06 March and
07 March, 2007 (HJD = 2454165.80 and HJD = 2454167.74)
respectively, at Sonoita Research Observatory (Sonoita, Arizona,
USA) using a robotic 0.35 meter telescope equipped with an SBIG
STL-1001XE CCD camera.

\IBVSfig{6cm}{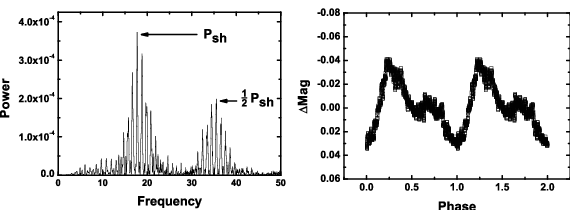}{\emph{Left:} Power spectrum, revealing
the $P_{sh}$ of SDSS J1021; \\ \emph{Right:} Superhump profile of
SDSS J1021} \IBVSfigKey{5763-f2.eps}{SDSS J102146.44+234926.3}{power
spectrum and phase diagram}

Spectroscopic observations were carried out on November 21.8 UT
with the CCD spectrograph mounted on the 1.01-m telescope of Bisei
Astronomical Observatory (Japan). The preliminary discussion of
the spectra can be found in (Ayani \& Kato, 2006). The spectral
range is 400-800nm, and the resolution is 0.5 nm at $H_{\alpha}$.
HR 3454 ($\alpha_{2000} = 08\hr 43\mm 13\fsec475; \delta_{2000} =
+03\deg 23\arcm 55\farcs18$) was observed for flux calibration of
the spectra. Standard IRAF routines were used for data reduction.

\begin{table}
\begin{center}
{\normalsize Table 2. Times of superhump maximums} \vskip 3mm
\begin{tabular}{cccc}
\hline HJD & E & O-C & $\sigma_{(O-C)}$\\
\hline

2454061.03380 & 0 & 0 & 0.00120\\

2454061.88103 & 15 & 0.00228  & 0.00130\\

2454061.93507 & 16 &  -0.00001  & 0.00368\\

2454061.99121 & 17 & -0.00020  & 0.00099\\

2454062.89325 & 33 & 0.00056 & 0.00179\\

2454062.94709 & 34 & -0.00193  & 0.00214\\

2454063.00533 & 35 & -0.00002  & 0.00156\\

2454063.62385 & 46 & -0.00113  & 0.00464\\

\hline
\end{tabular}
\end{center}
\end{table}

Spectrum (Fig. 3) shows blue continuum and Balmer absorption lines
(from $H_{\epsilon}$ to $H_{\alpha}$) together with K CaII 3934 in
absorption. Very weak HeI 4471, Fe 5169, NII 5767 absorption lines
may be present. $H_{\epsilon}$ 3970 is probably blended by H Ca II
3968. The FeIII 5461 line resembles weak P-Cygni profile.
Noteworthy, FeIII 5461 and NII 5767 may be artifacts caused by
imperfect subtraction of city lights: HgI 5461 and 5770 (spectrum
of the sky background which was subtracted, is available upon
request). The HeI 5876 line (mentioned for this object in Rau et
al., 2006) is not detectable on our spectrum.  It is remarkable
that $H_{\alpha}$ manifests a "W-like" profile: an emission
component embedded in the absorption component of the line.

Table~3 represents EWs (equivalent widths) of detected spectral
lines. EW was calculated by direct numerical integration over the
area under the line profile.

\begin{table}
\begin{center}
{\normalsize Table 3. Equivalent widths of spectral lines} \vskip
3mm
\begin{tabular}{cc}
\hline Line  & EW [\AA]\\
\hline
K CaII 3934 & -5.8\\
$H_{\epsilon}$ 3970 / H CaII 3968 & -8.7\\
$H_{\delta}$ 4101 & -6.4\\
$H_{\gamma}$ 4340 & -8.5\\
$H_{\beta}$ 4861 & -6.4\\
$H_{\alpha}$ 6563 & -7.7\\
$H_{\alpha}$ 6563 (emission) & 2.3\\
HeI 4471 & -0.95\\
FeII 5169 & -0.65\\
NII 5767 & -0.7\\
\hline
\end{tabular}
\end{center}
\end{table}

The archive photographic plates from the Main Astronomical
Observatory Wide Field Plate Archive (Kyiv, Ukraine; MAO
hereafter) and Plate Archive of Sternberg Astronomical Institute
of Moscow State University (Moscow, Russia; SAI hereafter) and
plate from Crimean Astrophysical Observatory archive (Ukraine)
were carefully scanned and inspected for previous outbursts on the
plates dating from 1978 to 1992 from MAO, 1913 - 1973 from SAI and
1948 from CrAO archives. The number of plates from each archive is
22 for SAI, 6 for MAO and 1 for CrAO archives. For all plates the
magnitude limit was determined (this data as well as scans of
plates are available upon request). The selection of plates from
MAO archive was done with the help of the database developed by
L.K. Pakuliak, which is accessible at
\emph{http://mao.kiev.ua/ardb/} (Sergeeva et al., 2004; Pakuliak,
L.K. \& Sergeeva, T.P., 2006;). No outbursts on the selected
plates from the MAO, SAI and CrAO archives were detected. This
implies that outbursts in SDDS J1021 are rather rare, which is
typical for the WZ Sge-type stars.

\IBVSfig{8cm}{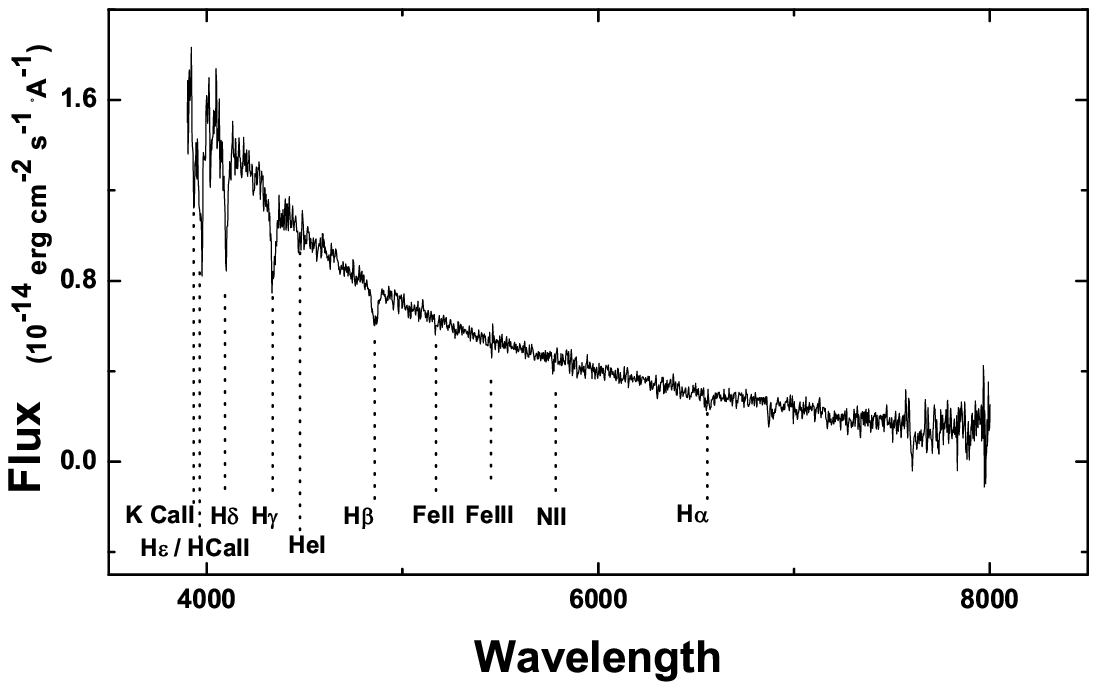}{Spectra of SDSS J1021 obtained on November
21.8 UT on 1.01-m telescope of Bisei Astronomical Observatory
(Japan)} \IBVSfigKey{5763-f3.eps}{SDSS J102146.44+234926.3}{Spectrum}

Table~4  (available only electronically from IBVS server or via
AAVSO ftp-server at
\emph{ftp://ftp.aavso.org/public/calib/varleo06.dat}) represents
$BVR_{c}I_{c}$ photometric calibration of 52 stars in SDSS J1021
vicinity, which have a V-magnitude in the range of
11\fmm21-17\fmm23 and can serve as a comparison stars. Calibration
(by AH$^8$) was done at Sonoita Research Observatory (Arizona,
USA).

The large amplitude of the SDSS J1021 outburst of 7\mm,
superhumps with a period below the ''period gap'',
rebrightening during the declining stage of superoutburst, rarity
of outbursts and obtained spectrum allow to classify this object
as a WZ Sge type dwarf nova.

\textbf{Acknowledgements:} AG is grateful to Aaron Price
(AAVSO, MA, USA) for his great help and useful discussions during
the preparation of this manuscript. Authors are thankful to A.
Zharova and L. Sat (both affiliated at SAI MSU, Moscow, RUSSIA)
for the assistance on dealing with SAI Plate Archive and to V.
Golovnya for the help concerning MAO Plate Archive (Kyiv,
Ukraine). It is a great pleasure to express gratefulness to Dr. N.
A. Katysheva, Dr. S. Yu. Shugarov (SAI MSU both) and Dr. D.
Bisikalo (Institute of Astronomy RAS, Moscow, Russia) for useful
discussions concerning the nature of SDSS J1021. IRAF is
distributed by the National Optical Astronomy Observatories, which
are operated by the Association of Universities for Research in
Astronomy, Inc., under cooperative agreement with the National
Science Foundation.

\IBVSedata{5763-t4.txt} 
\IBVSdataKey{5763-t4.txt}{SDSS J102146.44+234926.3}{photometric sequence}

\IBVSedata{5763-t5.txt}
\IBVSdataKey{5763-t5.txt}{SDSS J102146.44+234926.3}{photometry}
\IBVSedata{5763-t6.txt}
\IBVSdataKey{5763-t6.txt}{SDSS J102146.44+234926.3}{photometry}

\references

Adelman-McCarthy J. et al., 2007, submitted to {\it ApJ
Supplements}

Ayani, K. \& Kato, T., 2006, {\it CBET}, {\bf 753}, 1. Edited by
Green, D.W.E.

Bisikalo D.V. et al., 2006, {\it Chinese Journal of Astronomy and
Astrophysics, Supplement},  {\bf 6}, 159

Buat-Menard, V. \& Hameury, J.-M., 2002, {\it A\&A}, {\bf 386},
891

Christensen, E.J., 2006, {\it CBET}, {\bf 746}, 1. Edited by
Green, D.W.E.

Golovin A. et al., 2005, {\it IBVS} No.\ 5611

Imada A. et al., 2006, {\it PASJ}, {\bf 58}, L23

Kato, T., 2002, {\it PASJ}, {\bf 54}, L11

Lasota, J.P., Hameury, J.M., Hure, J.M., 1995, {\it A\&A}, {\bf
302}, L29

Marsakova V., Andronov, I.L., 1996, {\it Odessa Astronom. Publ.},
{\bf 9}, 127

Matthews, O.M. et al., 2005, {\it ASPC}, {\bf 330}, 171, 
in {\it The Astrophysics of Cataclysmic Variables and Related Objects, 
Eds. J.-M. Hameury and J.-P. Lasota. San Francisco: Astronomical Society 
of the Pacific

Osaki, Y., Meyer, F. \& Meyer-Hofmeister, E. 2001, {\it A\&A},
{\bf 370}, 488

Osaki, Y., \& Meyer, F., 2002, {\it A\&A}, {\bf 383}, 574

Osaki, Y., \& Meyer, F., 2003, {\it A\&A}, {\bf 401}, 325

Osaki, Y., 2003, {\it PASJ}, {\bf 55}, 841

Pakuliak, L.K. \& Sergeeva, T.P., 2006,  in {\it Virtual
Observatory: Plate Content Digitization, Archive Mining and Image
Sequence Processing}, Eds.: Tsvetkov, M., et al., Sofia, p.129

Patterson, J., et al., 1996, {\it PASP}, {\bf 108}, 748

Patterson, J., et al., 1998, {\it PASP}, {\bf 110}, 1290

Patterson, J., et al., 2002, {\it PASP}, {\bf 114}, 721

Pavlenko, E., et al., 2007, {\it In Proc. of the 15th European
White Dwarf Workshop "EUROWD06"}, in press.

Rau, A., et al., 2006, {\it The Astronomer's Telegram}, No.  951

Schreiber, M.R. \& Gansicke, B.T., 2001, {\it A\&A}, {\bf 375},
937

Sergeeva, T.P., et al., 2004, {\it Baltic Astronomy}, {\bf 13},
677

Templeton M. R. et al., 2006, {\it PASP}, {\bf 118}, 236

Waagen, Elizabeth O., 2006, {\it AAVSO Special Notice}, {\bf \#25}

\endreferences

\end{document}